\documentclass{aa}
\usepackage{psfig}
\usepackage{times}

\newcommand{\Hb}{\ifmmode {\rm H}\beta
\else H$\beta$\fi}

\newcommand{\Hp}{H$^{+}$}
\newcommand{\Hep}{He$^{+}$}
\newcommand{\Cppp}{C$^{+++}$}
\newcommand{\Nppp}{N$^{+++}$}
\newcommand{\Op}{O$^{+}$}
\newcommand{\Opp}{O$^{++}$}
\newcommand{\Oppp}{O$^{+++}$}

\newcommand{\oiii}{[O\,{\sc{iii}}]}

\newcommand{\Oiitonea}{[O\,{\sc{ii}}]\,$\lambda$3726,3729}
\newcommand{\Oiitoneb}{[O\,{\sc{ii}}]\,$\lambda$7320,7330}
\newcommand{\Oiitonec}{O\,{\sc{ii}}\,$\lambda$4651}

\newcommand{\Oiiitonea}{[O\,{\sc{iii}}]\,$\lambda$52\,$\mu$m}
\newcommand{\Oiiitoneb}{[O\,{\sc{iii}}]\,$\lambda$88\,$\mu$m}

\newcommand{\Ciii}{C\,{\sc{iii}}]\,$\lambda$1909}
\newcommand{\Nii}{[N\,{\sc{ii}}]\,$\lambda$6584}
\newcommand{\Oi}{[O\,{\sc{i}}]\,$\lambda$6300}
\newcommand{\Oii}{[O\,{\sc{ii}}]\,$\lambda$3727}
\newcommand{\Oiii}{[O\,{\sc{iii}}]\,$\lambda$5007}
\newcommand{\Oiiit}{[O\,{\sc{iii}}]\,$\lambda$4363}
\newcommand{\Neiii}{[Ne\,{\sc{iii}}]\,$\lambda$3869}

\newcommand{\rNii}{[N\,{\sc{ii}}]\,$\lambda$5755/6584}
\newcommand{\rOiii}{[O\,{\sc{iii}}]\,$\lambda$4363/5007}
\newcommand{\rSii}{[S\,{\sc{ii}}]\,$\lambda$6731/6717}
\newcommand{\rAriv}{[Ar\,{\sc{iv}}]\,$\lambda$4741/4713}

\newcommand{\qh}{$Q({\rm{H^{0}}})$}

\newcommand{\msun}{\ifmmode M_{\odot} \else M$_{\odot}$\fi}
\newcommand{\rsun}{\ifmmode R_{\odot} \else R$_{\odot}$\fi}
\newcommand{\lsun}{\ifmmode L_{\odot} \else L$_{\odot}$\fi}
\newcommand{\zsun}{\ifmmode Z_{\odot} \else $Z_{\odot}$\fi}

\newcommand{\Hepp}{He$^{++}$}

\begin{document}

\title{The temperature structure of dusty planetary nebulae}


\author{Gra\.{z}yna Stasi\'{n}ska \inst{1}, Ryszard Szczerba \inst{2}}

\institute{
            DAEC, Observatoire de Paris-Meudon, F-92195 Meudon Cedex, France,
            (grazyna.stasinska@obspm.fr)
\and
           N. Copernicus Astronomical Center, Rabia\'{n}ska 8,
           87--100 Toru\'{n}, Poland,
           (szczerba@ncac.torun.pl)
}

\date{received date; accepted date}

\titlerunning{The temperature in dusty planetary nebulae}
\authorrunning{Stasi\'{n}ska \& Szczerba}
\offprints{G.\ Stasi\'{n}ska}

\abstract{
We have analyzed the effects of photoelectric heating by dust grains
in photoionization models of planetary nebulae.
We have shown that this process is
particularly important if planetary nebulae
contain a population of small grains.
The presence of such grains would solve a
number of problems that have found no satisfactory solution so far: i) the
thermal energy deficit in some objects inferred from tailored
photoionization modelling; ii) the large negative temperature gradients
inferred directly from spatially resolved observations and indirectly from
integrated spectra in some planetary nebulae;
iii) the fact that the temperatures derived from the Balmer jump are
smaller than those derived from \rOiii;
iv) the fact that the observed intensities of \Oi\ are often larger than
predicted by photoionization models.
In presence of moderate density inhomogeneities,
such as inferred from high resolution images of planetary nebulae,
photoelectric heating would boost the temperature of the
tenuous component, which would then better confine the clumps.
The temperature structure of such
dusty and filamentary nebulae would solve the
long standing problem of temperature fluctuations posed by
Peimbert\,(\cite{peimbert67}).
}

\maketitle

\section{Introduction}

Infrared observations of planetary nebulae have established the presence
of dust in these objects (see e.g. reviews by Barlow\,\cite{barlow83} and
Roche\,\cite{roche89}).
Infrared spectra of planetary nebulae clearly show continuum emission 
in the mid- to
far-infrared range due to dust thermal emission (see e.g. Pottasch et
al.\,\cite{pottasch84}). Superimposed on this continuum are dust features
attributed to silicate dust particles (e.g. Aitken et al.\,\cite{aitken79},
Waters et al.\,\cite{waters98}, Cohen et al.\,\cite{cohen99}) or
carbon-based dust particles (e.g. Gillett et al.\cite{gillett73},
Waters et al.\,\cite{waters98}, Cohen et al.\,\cite{cohen99}, Szczerba et
al.\,\cite{szczerba01}).
The large depletion factors of
refractory elements such as Fe, Mg and Si (Shields\,\cite{shields78},
P\'equignot \& Stasi\'nska\,\cite{pequignot80}, Shields\,\cite{shields83},
Barlow\,\cite{barlow93}) and Ca (Aller \& Czyzak\,\cite{aller83},
Volk et al.\,\cite{volk97}) indicate that dust is intimately mixed 
with the ionized gas.

In a former paper (Stasi\'nska \& Szczerba{\,\cite{stasinska99}), we
performed a statistical analysis of a large sample
of planetary nebulae with available IRAS photometric data
by comparison with a grid of photoionization models for dusty planetary nebulae
in which we computed the infra-red emission of dust
heated by the absorbed stellar photons. The conclusion was that
planetary nebulae contain dust in a proportion from about $10^{-4}$ up to
$10^{-2}$ by mass (adopting the canonical grain size distribution
of Mathis et al.\,\cite{mathis77} -- hereafter the MRN distribution), and
that there is no sign of an evolution of the dust content on a time scale
of about $10^4$\,yrs, contrary to earlier suggestions (Natta \&
Panagia\,\cite{natta81}, Pottasch\,\cite{pottasch84}, Lenzuni et
al.\,\cite{lenzuni89}).

That dust in ionized plasmas plays an important role in the thermal
balance has been known since Spitzer\,(\cite{spitzer48}), who layed
the basis for a quantitative modelling of the effects of heating and
cooling of dusty plasmas.  Baldwin et al.\,(\cite{baldwin91}),
Borkowski \& Harrington\,(\cite{borkowski91}), Binette et
al.\,(\cite{binette93}), Viegas \& Contini\,(\cite{viegas94}) and
more recently Dopita \& Sutherland\,(\cite{dopita00}) have included the
effects of dust on the thermal balance of the gas in their photoionization
codes. Of these, only the studies of Borkowski \&
Harrington\,(\cite{borkowski91}) and Dopita \& Sutherland\,(\cite{dopita00})
deal explicitly with planetary nebulae. Borkowski \&
Harrington\,(\cite{borkowski91}) studied a very specific object, the galactic
halo planetary nebula IRAS\,18333$-$2357 located in the globular cluster M\,22,
which is hydrogen-poor and extremely dusty. They showed that in this
object,
the heating of the nebular gas is essentially provided by the photoelectric
emission from dust grains. Dopita \& Sutherland\,(\cite{dopita00})
studied photoelectric heating by dust in
normal planetary nebulae. They found its role to be important
especially if one assumes the presence of small grains in a much
higher proportion than inferred from the canonical MRN distribution.

In this paper, we develop further on the ideas of Dopita \&
Sutherland\,(\cite{dopita00}),
addressing the temperature distribution
inside planetary nebulae in more detail. This study is
motivated by the fact that there are a number of unsolved
problems relating to the temperature structure of planetary nebulae.
The best known
concerns the temperature fluctuations, whose
existence has been postulated by Peimbert\,(\cite{peimbert67}) to explain the
difference between the temperature $T_{\rm r[O\,III]}$
derived from the \rOiii\ line ratio
and the one derived from the Balmer jump. Using Peimbert's formulation,
one derives a temperature fluctuation $t^{2}$ typically of 0.03--0.04
in planetary nebulae (Liu \& Danziger\,\cite{liu93}, see also
Esteban\, \cite{esteban01} for a recent review).
Although the Peimbert's approach relies
on simplifications which are not necessarily justified (e.g. Kingdon
\& Ferland\,\cite{kingdon95}) and the real temperature variations may not be
adequately described by his scheme (Stasi\'nska\,\cite{stasinska01}), the
mere fact that the Balmer jump temperature is significantly lower than
$T_{\rm r[O\,III]}$ in a large number of planetary nebulae indicates
that their temperature is far from uniform. While
photoionization models do predict temperature
gradients in planetary nebulae, as a result of the gradual modification of the
ionizing radiation field and of the radial distributions of the ions 
responsible
for the heating and the cooling of the gas, the computed gradient is
too small to account for the difference between the  Balmer jump temperature
and
$T_{\rm r[O\,III]}$. Moderate density fluctuations are unable to
produce temperature fluctuations on a  $t^{2}$ = 0.03 level (Kingdon
\& Ferland\,\cite{kingdon95}).
Very high density clumps ($n\sim10^6$\,cm$^{-3}$)
can bias upwards the temperature derived from the \rOiii\
ratio due to collisional deexcitation of the \Oiii\ line
(Viegas \& Clegg\,\cite{viegas94a}).
However, the existence of such high densities in the  \Opp\
emitting zone is not supported by the
density sensitive line ratios observed in planetary nebulae both from
the infrared fine structure  \oiii\ lines and from the \rAriv\ doublet
(e.g. Liu et al.\,\cite{liu01a}). The magnitude of the temperature gradient
observed in planetary nebulae either indirectly from a comparison of
  $T_{\rm r[O\,III]}$ with $T_{\rm r[N\,II]}$, the
temperature derived from \rNii\,
(see e.g. Kingsburgh \&
Barlow\,\cite{kingsburgh94} or Pe\~na et al.\,\cite{pena01})
or directly from spatially resolved observations
(Liu et al.\,\cite{liu00},
Garnett \& Dinerstein\,\cite{garnett01}) is often difficult to
account for by photoionization models.

In this paper, we explore whether photoionization models
including the effects of dust on the thermal balance of the emitting
gas can answer some of these questions.

In the next section, we briefly describe the photoionization code and
dust properties. In Section 3, we study the temperature profiles of
planetary nebulae with smooth density distributions. In Section 4, we discuss
the temperature profiles of planetary nebulae with
small scale density inhomogeneities. In Section
5, we discuss the effects of dust on methods of analysis of planetary nebulae
based on emission line ratios. In Section 6 we discuss the
temperature fluctuation parameter. Concluding remarks are given in Section 7.

\section{The building of dusty photoionization models}

\subsection{The treatment of dust in the photoionization code}

The present paper is based on results from the photoionization code PHOTO
using the same atomic data as Stasi\'nska \& Leitherer\,(\cite{stasinska96}).
All the computations have been done under the assumption of spherical symmetry
and adopting the outward only approximation for the diffuse
ionizing radiation.

In the version of PHOTO used by Stasi\'nska \& Szczerba\,(\cite{stasinska99}),
we had adopted the formulation of Harrington et al.\,(\cite{harrington88})
to compute the absorption of photons by dust grains,
and their conversion into thermal radiation from the grains.
For the present paper, we have also incorporated the effects of grains on the
thermal balance of the gas, by implementing the
physical processes between dust, gas and radiation relevant for dusty
ionized regions as described in Appendix C of Baldwin et
al.\,(\cite{baldwin91}) -- except that we do not consider an average grain
size like former authors, but consider all the particle sizes explicitly.
  In short, the steady state grain potential is
determined self-consistently for each dust particle size and the corresponding
heating of the electron plasma by the photoelectric emission from dust
is computed.
As a check, the same treatment of dust has been applied to the
photoionization code  written by Szczerba (see e.g. G{\c{e}}sicki et
al.\,\cite{gesicki96}). Both codes give consistent results and
reproduce the Baldwin et
al.\,(\cite{baldwin91}) model for the Orion Nebula satisfactorily
(their Fig.\,16).

For simplicity, in this paper we consider only graphite grains
(Draine \& Laor\,\cite{draine93}) in our planetary nebulae models.
We include two population of grains. One is composed of graphite grains with
a bulk density of
2.5\,g\,cm$^{-3}$ and sizes distributed between 0.01
and 0.25\,$\mu$m according to the classical MRN model
(hereafter referred to as standard grains).
  The second one consists of small grains, as
introduced by Dopita \& Sutherland\,(\cite{dopita00}).
We assume that the small grain have a  bulk density of
1\,g\,cm$^{-3}$ and that their sizes are distributed according to the
same MRN law but between
0.001 and 0.01\,$\mu$m.
  The formulae describing the photoelectron energy
distribution and the yield are from Dopita \& Sutherland\,(\cite{dopita00}).
As explained by these authors, in comparison with the
formulation of Baldwin et al.\,(\cite{baldwin91})
the adopted approach seems to be more realistic.

\subsection{The planetary nebula models}

We consider two families of photoionization models for
planetary nebulae.
The first one corresponds to a rather dense nebula ionized by a star that is
not very hot. The reference model has a density
$n(\rm H)$\,=\,10$^{4}$ cm$^{-3}$
and a stellar temperature $T_*$\,=\,5\,10$^4$\,K. The second one
corresponds to a more diluted nebula ionized by a hotter star. We have
chosen for the reference model $n(\rm H)$\,=\,5\,10$^{3}$ cm$^{-3}$
and $T_*$\,=\,10$^5$\,K. The stars are assumed to radiate as blackbodies.
The inner radius of the nebula is chosen to be 5\,10$^{16}$\,cm for
the former case, and 8\,10$^{16}$ cm for the latter one.
In both cases, we assume that the total
number of H Lyman continuum photons emitted by the star,
\qh\,  is equal to 3\,10$^{47}$ ph\,s$^{-1}$ and that the
nebulae are ionization
bounded (the calculations are stopped when the proportion of neutral hydrogen
exceeds 0.03). This implies that the total nebular mass is larger
than $\simeq$ 0.1\,$M_\odot$ for the first case, and larger than
  $\simeq$ 0.3\,$M_\odot$ for the second
case, which are reasonable values. Actually, the local properties of our
models do not depend
on whether the nebula is ionization- or density-bounded, but the
integrated properties do depend on that, since the low excitation
parts are suppressed for density-bounded nebulae.
In Table\,1 and Table\,2 we have collected the key parameters used in the
models computed by us for $T_*$\,=\,5\,10$^4$\,K and 10$^5$\,K, respectively.
These tables allow to easily relate the name of the model to the physical
parameters adopted in that model.

\begin{table*}
\caption{Definition of the models with $T_*$\,=\,5\,10$^4$\,K.}
\begin{flushleft}
\begin{tabular}{lccccccc}
\hline
  model    &
   Aa &
   Ba &
   Ca &
   Da &
   Ea &
   Fa &
   Ga \\
\hline
$T_*$ [K]&  5\,10$^4$  &  5\,10$^4$  &  5\,10$^4$   &  5\,10$^4$  & 
5\,10$^4$   &  5\,10$^4$   &  5\,10$^4$ \\
$Q({\rm{H^{0}}})$ [ph\,s$^{-1}$]  & 3\,10$^{47}$  & 3\,10$^{47}$  & 
3\,10$^{47}$  & 3\,10$^{47}$ & 3\,10$^{47}$ & 3\,10$^{47}$ & 
3\,10$^{47}$ \\
$R_{\rm in}$ [cm] &  5\,10$^{16}$ & 5\,10$^{16}$ & 5\,10$^{16}$ & 
5\,10$^{16}$ & 5\,10$^{16}$ & 5\,10$^{16}$ & 5\,10$^{16}$ \\
   &   &   &   &   &   &   \\
$n(\rm H)$ [cm$^{-3}$]  & $10^{4}$ & $10^{4}$ & $10^{4}$ & $10^{4}$ & 
Eq (3) & Eq (4)  & Eq (4) \\
   &   &   &   &   &   &   \\
dust properties  & no grains &  standard grains & standard +      & 
standard +  & no grains  & no grains & standard +  \\
                  &           &            & small grains    & small 
grains &           &           & small grains \\
$m_{\rm d}/m_{\rm H}$ &  0. & 10$^{-2}$ & 5\,10$^{-3}$+5\,10$^{-3}$ 
& 5\,10$^{-4}$+5\,10$^{-4}$   & 0. &  0. &  5\,10$^{-3}$+5\,10$^{-3}$ 
\\
\hline
\end{tabular}
\end{flushleft}
\label{Table1}
\end{table*}
\begin{table*}
\caption{Definition of the models with $T_*$\,=\,10$^5$\,K.}
\begin{flushleft}
\begin{tabular}{lccccccc}
\hline
model   &
  Ab &
  Bb &
  Cb &
  Db &
  Eb &
  Fb &
  Gb \\
\hline
$T_*$ [K]&  10$^5$  &  10$^5$  &  10$^5$   &  10$^5$  &  10$^5$   & 
10$^5$   &  10$^5$\\
$Q({\rm{H^{0}}})$ [ph\,s$^{-1}$] & 3\,10$^{47}$  & 3\,10$^{47}$  & 
3\,10$^{47}$  & 3\,10$^{47}$ & 3\,10$^{47}$ & 3\,10$^{47}$ & 
3\,10$^{47}$ \\
$R_{\rm in}$ [cm] &  8\,10$^{16}$ & 8\,10$^{16}$ & 8\,10$^{16}$ & 
8\,10$^{16}$ & 8\,10$^{16}$ & 8\,10$^{16}$ & 8\,10$^{16}$ \\
   &   &   &   &   &   &   \\
$n(\rm H)$ [cm$^{-3}$]  & $5\,10^{3}$ & $5\,10^{3}$ & $5\,10^{3}$ & 
$5\,10^{3}$ & Eq (3) & Eq (4)  & Eq (4) \\
   &   &   &   &   &   &   \\
dust properties  & no grains &  standard grains & standard +      & 
standard +  & no grains  & no grains & standard +  \\
                  &           &            & small grains    & small 
grains &           &           & small grains \\
$m_{\rm d}/m_{\rm H}$ &  0. & 10$^{-2}$ & 5\,10$^{-3}$+5\,10$^{-3}$ 
& 5\,10$^{-4}$+5\,10$^{-4}$   & 0. &  0. &  5\,10$^{-3}$+5\,10$^{-3}$ 
\\
\hline
\end{tabular}
\end{flushleft}
\label{Table2}
\end{table*}

Our models have the following chemical composition:
He/H = 0.115; C/H = 5.50\,10$^{-4}$; N/H = 2.24\,$10^{-4}$;
O/H = 4.79\,$10^{-4}$; Ne/H = 1.23\,$10^{-4}$; Mg/H = 3.80\,$10^{-5}$;
Si/H = 3.55\,$10^{-5}$; S/H = 8.32\,$10^{-6}$; Cl/H = 1.86\,$10^{-7}$;
Ar/H = 2.46\,$10^{-6}$; Fe/H = 4.68\,$10^{-5}$,
which corresponds to the average chemical composition of the Galactic
planetary nebulae from the Kingsburgh \& Barlow\,(\cite{kingsburgh94})
sample (Mg, Si, Cl and Fe have
solar abundances). This chemical
composition is assumed throughout or study, regardless of the dust
abundance. The reason for adopting such a policy is its simplicity.
Of course, if metallic or carbon atoms are condensed into grains, their gas
phase abundances are correspondingly depleted. However, such elements
as Si, Mg, Fe do not play a dominant role in the energy balance.
Carbon, on the other hand, can play an important role, especially
in carbon-rich planetary nebulae. However, we are not interested
here to investigate what would happen to a planetary nebula of given
chemical composition if part of its carbon content were locked into
grains. We rather wish to explore what are the effects of dust on the
temperature structure of planetary nebulae.

\begin{table*}
\caption{Selected properties for models with $T_*$\,=\,5\,10$^4$\,K.}
\begin{flushleft}
\begin{tabular}{lccccccc}
\hline
    &
  model Aa &
  model Ba &
  model Ca &
  model Da &
  model Ea &
  model Fa &
  model Ga \\
\hline
$F$(H$\beta$)                        &    1.149E-09 & 8.082E-10 & 
1.774E-10 & 7.250E-10 & 1.137E-09 & 1.145E-09 & 2.202E-10 \\
   &   &   &   &   &   &   \\
$F_{12}^{\rm IRAS}$                  &	 6.175E+00 & 2.819E+01 & 
1.420E+02 & 5.914E+01 & 5.155E+00 & 5.342E+00 & 1.400E+02 \\
$F_{25}^{\rm IRAS}$                  &	 5.667E+00 & 3.885E+02 & 
6.182E+02 & 2.484E+02 & 4.572E+00 & 4.603E+00 & 5.926E+02 \\
$F_{60}^{\rm IRAS}$                  &	 1.473E+01 & 3.865E+02 & 
3.445E+02 & 1.376E+02 & 1.048E+01 & 1.007E+01 & 3.154E+02 \\
$F_{100}^{\rm IRAS}$                 &	 3.302E+00 & 7.361E+01 & 
5.499E+01 & 2.173E+01 & 2.200E+00 & 2.290E+00 & 4.912E+01 \\
   &   &   &   &   &   &   \\
He\,{\sc{i}}\,$\lambda$5876          &    1.680E-01 & 1.711E-01 & 
1.890E-01 & 1.745E-01 & 1.693E-01 & 1.689E-01 & 1.882E-01 \\
He\,{\sc{ii}}\,$\lambda$4686         &    1.728E-03 & 2.405E-03 & 
8.745E-03 & 2.556E-03 & 1.763E-03 & 1.730E-03 & 7.367E-03 \\
   &   &   &   &   &   &   \\
\Oi                              &    6.771E-03 & 8.911E-03 & 
6.110E-02 & 1.615E-02 & 2.651E-03 & 1.350E-02 & 7.165E-02 \\
\Oiitonea                            &    5.428E-01 & 6.018E-01 & 
1.466E+00 & 7.069E-01 & 4.573E-01 & 4.612E-01 & 1.187E+00 \\
\Oiitoneb                            &    6.371E-02 & 7.428E-02 & 
2.564E-01 & 9.629E-02 & 6.211E-02 & 1.053E-01 & 3.362E-01 \\
\Oiitonec                            &    4.101E-03 & 4.165E-03 & 
4.302E-03 & 4.396E-03 & 4.450E-03 & 3.649E-03 & 3.779E-03 \\
O\,{\sc{iii}}]\,$\lambda$1663        &    9.149E-03 & 1.385E-02 & 
6.448E-01 & 3.619E-02 & 1.134E-02 & 9.393E-03 & 4.560E-01 \\
\Oiiit                               &    1.530E-02 & 2.067E-02 & 
2.942E-01 & 4.051E-02 & 1.870E-02 & 1.553E-02 & 2.148E-01 \\
\Oiii                                &    4.627E+00 & 5.410E+00 & 
1.882E+01 & 7.625E+00 & 5.217E+00 & 4.264E+00 & 1.461E+01 \\
\Oiiitonea                           &    3.486E-01 & 3.650E-01 & 
5.100E-01 & 4.107E-01 & 2.502E-01 & 2.380E-01 & 3.276E-01 \\
\Oiiitoneb                           &    4.392E-02 & 4.603E-02 & 
6.481E-02 & 5.189E-02 & 2.936E-02 & 3.058E-02 & 4.259E-02 \\
   &   &   &   &   &   &   \\
\Ciii                                &    3.095E-01 & 4.357E-01 & 
9.567E+00 & 9.298E-01 & 3.663E-01 & 3.244E-01 & 6.858E+00 \\
   &   &   &   &   &   &   \\
\Nii                                 &    1.334E+00 & 1.391E+00 & 
2.514E+00 & 1.466E+00 & 1.071E+00 & 1.589E+00 & 2.790E+00 \\
   &   &   &   &   &   &   \\
\Neiii                               &    3.679E-01 & 4.486E-01 & 
2.119E+00 & 6.829E-01 & 4.192E-01 & 3.584E-01 & 1.689E+00 \\
   &   &   &   &   &   &   \\
\rOiii                               &    3.308E-03 & 3.822E-03 & 
1.564E-02 & 5.313E-03 & 3.585E-03 & 3.643E-03 & 1.470E-02 \\
\rNii                                &    1.132E-02 & 1.215E-02 & 
1.940E-02 & 1.407E-02 & 1.248E-02 & 1.752E-02 & 2.535E-02 \\
($F\lambda$3642-$F\lambda$3648)/$F$(\Hb)  &    4.902E-03 & 4.801E-03 
& 4.026E-03 & 4.593E-03 & 4.852E-03 & 4.855E-03 & 4.132E-03 \\
\rSii                                &    1.991E+00 & 1.980E+00 & 
1.860E+00 & 1.924E+00 & 1.896E+00 & 2.097E+00 & 1.986E+00 \\
\rAriv                               &    1.691E+00 & 1.673E+00 & 
1.500E+00 & 1.629E+00 & 2.340E+00 & 1.885E+00 & 1.598E+00 \\
   &   &   &   &   &   &   \\
$T_0$(H$^+$)                           &    8151.	    & 8427 
& 11807.    & 9070.     & 8279.     & 8268.	& 11208.    \\
$T_0$(O$^+$)                           &    7972.	    & 8184. 
& 10010.    & 8672.     & 8116.	    & 8162.     & 9870.     \\
$T_0$(O$^{++}$)                        &    8232.	    & 8529. 
	& 12360.    & 9204.     & 8334.	    & 8335.     & 11836.    \\
   &   &   &   &   &   &   \\
$t^{2}$(\Hp)                         &     8.0E-4   &   1.3E-3  & 
4.6E-2  &   4.5E-3  &   7.0E-4  &   6.6E-4  &   4.4E-2  \\
$t^{2}$(\Op)                         &     1.3E-3   &   1.6E-3  & 
1.2E-2  &   1.9E-3  &   1.6E-4  &   7.7E-4  &   7.3E-3  \\
$t^{2}$(\Opp)                        &     2.9E-4   &   6.7E-4  & 
4.3E-2  &   4.4E-3  &   2.4E-4  &   4.3E-4  &   4.6E-2  \\
\hline
\end{tabular}
\end{flushleft}
\label{Table3}
\end{table*}
\begin{table*}
\caption{Selected properties of models with $T_*$\,=\,10$^5$\,K.}
\begin{flushleft}
\begin{tabular}{lccccccc}
\hline
    &
  model Ab &
  model Bb &
  model Cb &
  model Db &
  model Eb &
  model Fb &
  model Gb \\
\hline
$F$(H$\beta$)                        & 1.185E-09 & 8.961E-10 & 
3.204E-10 & 8.930E-10 & 1.163E-09 & 1.169E-09	 & 2.939E-10 \\
   &   &   &   &   &   &   \\
$F_{12}^{\rm IRAS}$                  & 1.064E+01 & 1.162E+01 & 
4.887E+01 & 2.260E+01 & 1.057E+01 & 8.337E+00	 & 8.595E+01 \\
$F_{25}^{\rm IRAS}$                  & 1.879E+01 & 2.224E+02 & 
5.616E+02 & 1.917E+02 & 1.193E+01 & 1.524E+01	 & 5.785E+02 \\
$F_{60}^{\rm IRAS}$                  & 3.721E+01 & 4.668E+02 & 
5.697E+02 & 1.936E+02 & 3.154E+01 & 2.452E+01	 & 4.604E+02 \\
$F_{100}^{\rm IRAS}$                 & 1.061E+01 & 1.199E+02 & 
1.185E+02 & 4.045E+01 & 8.651E+00 & 6.967E+00	 & 8.842E+01 \\
   &   &   &   &   &   &   \\
He\,{\sc{i}}\,$\lambda$5876          & 1.636E-01 & 1.606E-01 & 
1.432E-01 & 1.616E-01 & 1.635E-01 & 1.672E-01	 & 1.471E-01 \\
He\,{\sc{ii}}\,$\lambda$4686         & 1.082E-01 & 1.362E-01 & 
3.212E-01 & 1.385E-01 & 1.123E-01 & 1.109E-01	 & 3.363E-01 \\
   &   &   &   &   &   &   \\
\Oi                              & 5.159E-02 & 5.559E-02 & 1.044E-01 
& 6.262E-02 & 2.313E-02 & 7.552E-02	 & 1.615E-01 \\
\Oiitonea                            & 1.110E+00 & 1.189E+00 & 
1.982E+00 & 1.306E+00 & 8.950E-01 & 1.022E+00	 & 1.757E+00 \\
\Oiitoneb                            & 1.086E-01 & 1.153E-01 & 
2.083E-01 & 1.315E-01 & 7.963E-02 & 1.821E-01	 & 3.282E-01 \\
\Oiitonec                            & 4.830E-03 & 4.670E-03 & 
3.905E-03 & 4.677E-03 & 5.192E-03 & 4.557E-03	 & 3.638E-03 \\
O\,{\sc{iii}}]\,$\lambda$1663        & 1.059E-01 & 1.206E-01 & 
4.862E-01 & 1.638E-01 & 1.365E-01 & 1.043E-01	 & 4.718E-01 \\
\Oiiit                               & 9.251E-02 & 1.005E-01 & 
2.578E-01 & 1.262E-01 & 1.144E-01 & 9.109E-02	 & 2.475E-01 \\
\Oiii                                & 1.255E+01 & 1.284E+01 & 
1.859E+01 & 1.445E+01 & 1.439E+01 & 1.189E+01	 & 1.727E+01 \\
\Oiiitonea                           & 8.893E-01 & 8.716E-01 & 
8.428E-01 & 9.008E-01 & 7.739E-01 & 5.950E-01	 & 5.527E-01 \\
\Oiiitoneb                           & 1.379E-01 & 1.353E-01 & 
1.321E-01 & 1.403E-01 & 1.143E-01 & 9.172E-02	 & 8.593E-02 \\
   &   &   &   &   &   &   \\
\Ciii                                & 2.093E+00 & 2.289E+00 & 
6.975E+00 & 2.931E+00 & 2.579E+00 & 2.095E+00	 & 6.675E+00 \\
   &   &   &   &   &   &   \\
\Nii                                 & 1.829E+00 & 1.955E+00 & 
3.042E+00 & 2.086E+00 & 1.217E+00 & 2.272E+00	 & 3.632E+00 \\
   &   &   &   &   &   &   \\
\Neiii                               & 1.341E+00 & 1.403E+00 & 
2.355E+00 & 1.609E+00 & 1.482E+00 & 1.353E+00	 & 2.307E+00 \\
   &   &   &   &   &   &   \\
\rOiii                               & 7.374E-03 & 7.828E-03 & 
1.387E-02 & 8.729E-03 & 7.948E-03 & 7.660E-03	 & 1.433E-02 \\
\rNii                                & 1.688E-02 & 1.668E-02 & 
1.877E-02 & 1.764E-02 & 1.826E-02 & 2.123E-02	 & 2.311E-02 \\
($F\lambda$3642-$F\lambda$3648)/$F$(\Hb)  & 4.279E-03 & 4.256E-03 & 
3.965E-03 & 4.165E-03 & 4.228E-03 & 4.269E-03	 & 3.979E-03 \\
\rSii                                & 1.690E+00 & 1.689E+00 & 
1.674E+00 & 1.685E+00 & 1.446E+00 & 1.809E+00	 & 1.828E+00 \\
\rAriv                               & 1.183E+00 & 1.179E+00 & 
1.147E+00 & 1.172E+00 & 1.430E+00 & 1.435E+00	 & 1.406E+00 \\
   &   &   &   &   &   &   \\
$T_0$(H$^+$)                           & 10410.	 & 10602.    & 12906. 
	 & 10988.    & 10615.	 & 10462.        & 13120.    \\
$T_0$(O$^+$)                           & 10097.	 & 10045.    & 10575 
& 10312.    & 10566.    & 10164.        & 10661.    \\
$T_0$(O$^{++}$)                        & 10310.    & 10489.    & 
12576.    & 10874.    & 10523.	 & 10347.        & 12597.    \\
   &   &   &   &   &   &   \\
$t^{2}$(\Hp)                         &     4.9E-3   &   6.7E-3  & 
3.4E-2  &   7.3E-3  &   3.7E-3  &   5.9E-3  &   6.5E-2  \\
$t^{2}$(\Op)                         &     2.4E-3   &   2.9E-3  & 
8.1E-3  &   3.2E-3  &   2.7E-3  &   1.3E-3  &   7.0E-3  \\
$t^{2}$(\Opp)                        &     2.1E-3   &   2.6E-3  & 
1.4E-2  &   2.7E-3  &   2.4E-3  &   2.6E-3  &   1.6E-2  \\
\hline
\end{tabular}
\end{flushleft}
\label{Table4}
\end{table*}

In Tables \ref{Table3} and \ref{Table4} we summarize some interesting 
features of the
models discussed in this paper. We list the total \Hb\ fluxes
in erg\,cm$^{-2}$\,s$^{-1}$ and the fluxes in the four IRAS bands at
12, 25, 60 and 100\,$\mu$m,
assuming that the nebulae lie at a distance of 1\,kpc. Below we give
the intensities relative to \Hb\ of helium and oxygen lines. These
are sufficient to give a fair understanding of the behaviour of the
emission line spectra.
We also list the intensities of the most
important lines of C, N and Ne used in abundance determinations.
We do not compute the reddening  by internal dust,
so that the listed line intensities are relevant
to observations corrected for reddening
(the dereddening procedure, ideally consisting in fitting the observed Balmer
decrement to the theoretical one,
  actually produces a small error in the ratios of lines arising from 
different ions).
Then we list the two temperature sensitive line
ratios \rOiii\ and \rNii, and the Balmer jump defined as
($F\lambda$3642-$F\lambda$3648)/$F$(\Hb), where $F\lambda$3642 and
$F\lambda$3648  are the nebular continuum
fluxes at 3642 and 3648 \AA. Below are given the two density sensitive
line ratios \rSii\ and \rAriv.

Next we give the values of
the
average ionic temperatures $T_0$(\Hp), $T_0$(\Op) and $T_0$(\Opp), where
  $T_0({\rm X}^{+j}$) is defined as:
\begin{equation}
T_0({\rm X}^{+j})=\frac
              {\int{T_{\rm e}\,n({\rm X}^{+j})\,n_{\rm e}{\rm d}V}}
              {\int{n({\rm X}^{+j})n_{\rm e}\,{\rm d}V}},
\end{equation}
and $T_{\rm e}$ is the local electron temperature, $n_{\rm e}$ the
electron density,
$n({\rm X}^{+j})$ stands for the ion concentration, the integration
being performed over the total volume.

The last three lines give the values of $t{^2}$(\Hp), $t{^2}$(\Op) 
and $t{^2}$(\Opp),
where  $t{^2}({\rm X}^{+j})$ is defined as:

\begin{equation}
t{^2}({\rm X}^{+j})=\frac
              {\int{(T_{\rm e}-T_0({\rm X}^{+j}))^2\,n({\rm 
X}^{+j})\,n_{\rm e}{\rm d}V}}
              {T_0({\rm X}^{+j})^2\,\int{n({\rm X}^{+j})n_{\rm e}\,{\rm d}V}}.
\end{equation}

\section{The effect of dust on the temperature profiles of planetary
nebulae with smooth density distributions}

\begin{figure*}
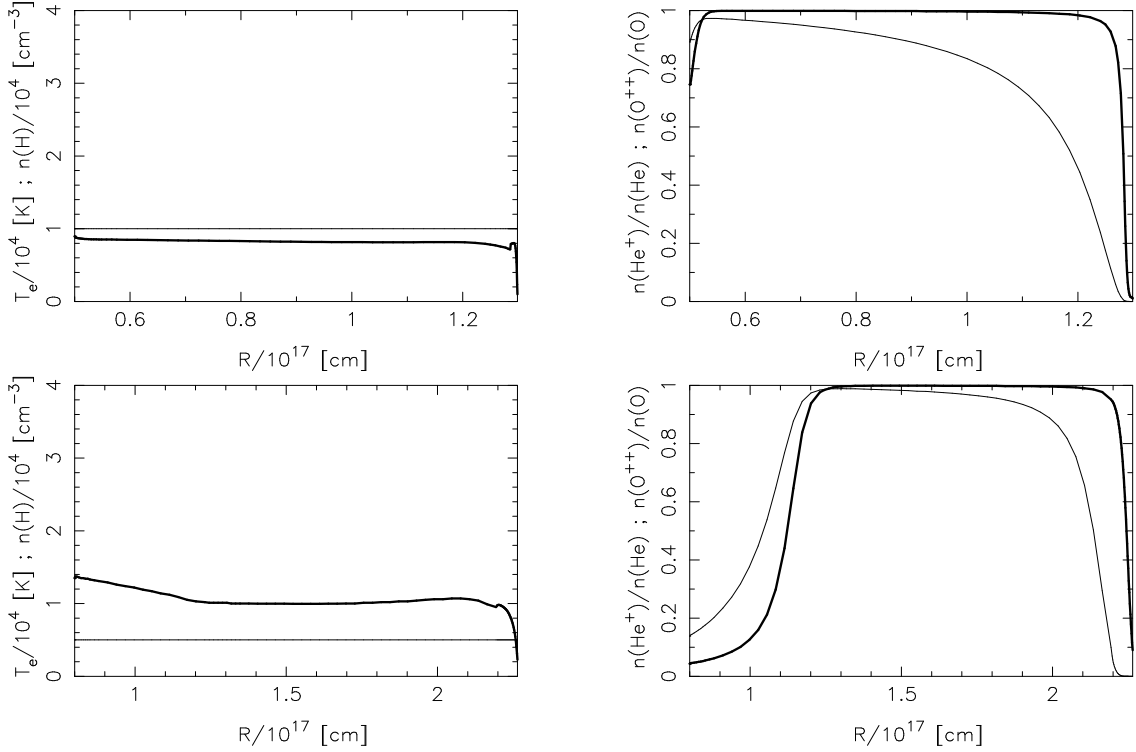

\centerline{\psfig{figure=MS1707f1a.ps,width=15.cm}}
\centerline{\psfig{figure=MS1707f1b.ps,width=15.cm}}
\caption{
The temperature and ionization structure of dust-free and constant
density model planetary nebulae (Models Aa and Ab). Upper panels: models with
$n({\rm H})$\,=\,10$^{4}$\,cm$^{-3}$ and $T_*$\,=\,5\,10$^4$\,K. Lower panels,
models with  $n({\rm H})$\,=5\,10$^{3}$ cm$^{-3}$ and $T_*$\,=\,10$^5$\,K.
Left: electron temperature (thick line) and hydrogen concentration
(thin line) as a function of the distance to the star; right: ionization
fraction of \Hep\ (thick line) and \Opp\ (thin line). Note that here
as well as in the other figures of the paper, the
horizontal scale matches the size of the ionized region and is
therefore different for each model, while the vertical scale is always
the same.
}
\label{Fig1}
\end{figure*}
We first investigate the simplest case of a uniform density distribution.
Figure\,\ref{Fig1} shows the results of dust-free models (Models Aa 
and Ab) for the
$n(\rm H)$\,=\,10$^4$ cm$^{-3}$,  $T_*$\,=\,5\,10$^4$\,K case (upper panels)
and for the  $n({\rm H})$\,=\,5\,10$^{3}$ cm$^{-3}$, $T_*$\,=\,10$^5$\,K case
(lower panels). The left panels show the electron temperature (thick line) and
hydrogen number density (thin line) as a function of the distance $R$ to the
star. The right panels show the ionization fraction of \Hep\ (thick line) and
\Opp\ (thin line).

In Figure\,\ref{Fig2}, we display with the same conventions the results of
dusty models (Ba and Bb) in which dust is composed of graphite grains with the
MRN size distribution between 0.01 and 0.25\,$\mu$m and a
dust-to-hydrogen mass ratio, $m_{\rm d}/m_{\rm H}$, of 10$^{-2}$.
\begin{figure*}
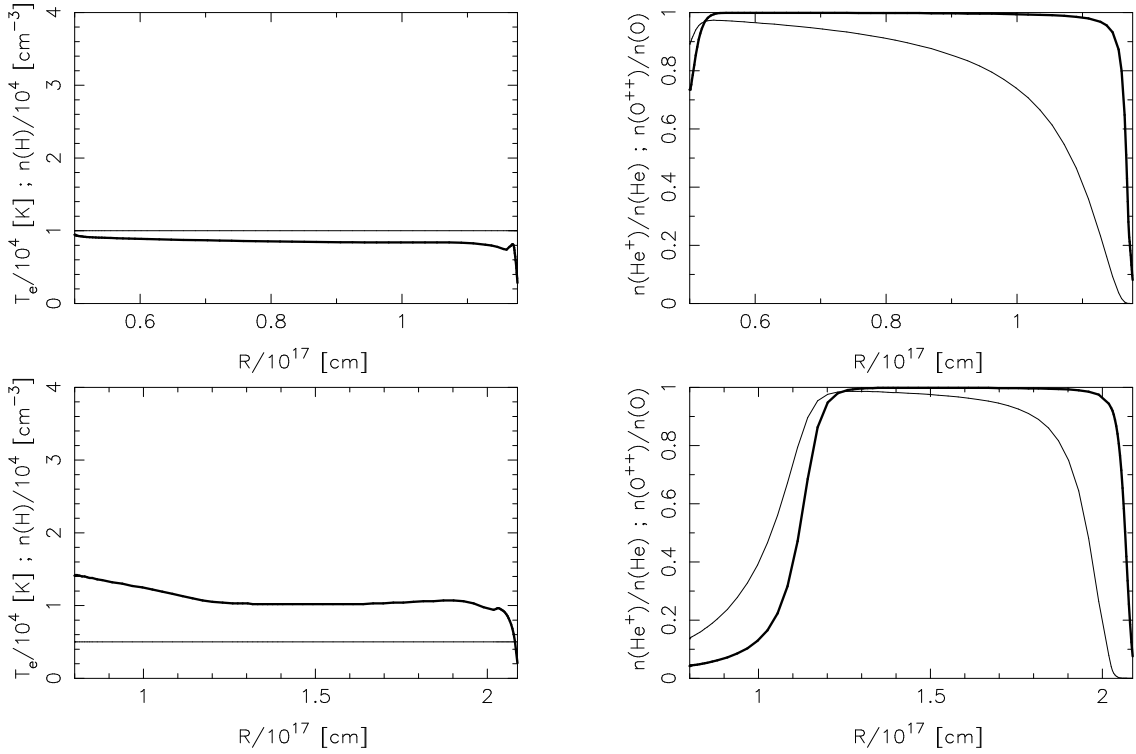

\centerline{\psfig{figure=MS1707f2a.ps,width=15.cm}}
\centerline{\psfig{figure=MS1707f2b.ps,width=15.cm}}
\caption{
The temperature and ionization structure of the same constant
density models as in Fig.\,\ref{Fig1} but containing graphite
grains with the MRN size distribution between 0.01 and 0.25 $\mu$m and a
dust-to-hydrogen mass ratio of 10$^{-2}$ (Models Ba and Bb). The
presentation is the same as for the models shown in Fig.\,\ref{Fig1}.
}
\label{Fig2}
\end{figure*}
Clearly, the size of the ionized zone is shrunken with respect to the
dust-free models shown in Fig.\,\ref{Fig1}, because with such
$m_{\rm d}/m_{\rm H}$ the grains
significantly compete with gas for the absorption of the Lyman continuum
photons. However, there is no strong effect neither on the temperature
structure of the nebula nor on the ratios of emission lines with respect to
\Hb, as can be seen in Table\,\ref{Table3} and \ref{Table4}.
The only notable difference is in the \Hb\ and the IRAS
fluxes. The \Hb\ flux is reduced in the dusty models by about 30\%, while
the fluxes in the IRAS bands are increased by large amounts due to
the thermal emission of the dust grains that have been heated by the
stellar and nebular photons (note that in the dust-free models, the fluxes
in the IRAS bands are not zero due to the contribution of the fine structure
lines from atomic ions).

Figure\,\ref{Fig3} shows models Ca and Cb which differ from models Ba and Bb
only in that half of the dust mass is now in small grains of size from 0.001
to 0.01\,$\mu$m. That is to say, large grains and small grains coexist, with
a dust-to-hydrogen mass ratio of 5\,10$^{-3}$ for each component.
\begin{figure*}
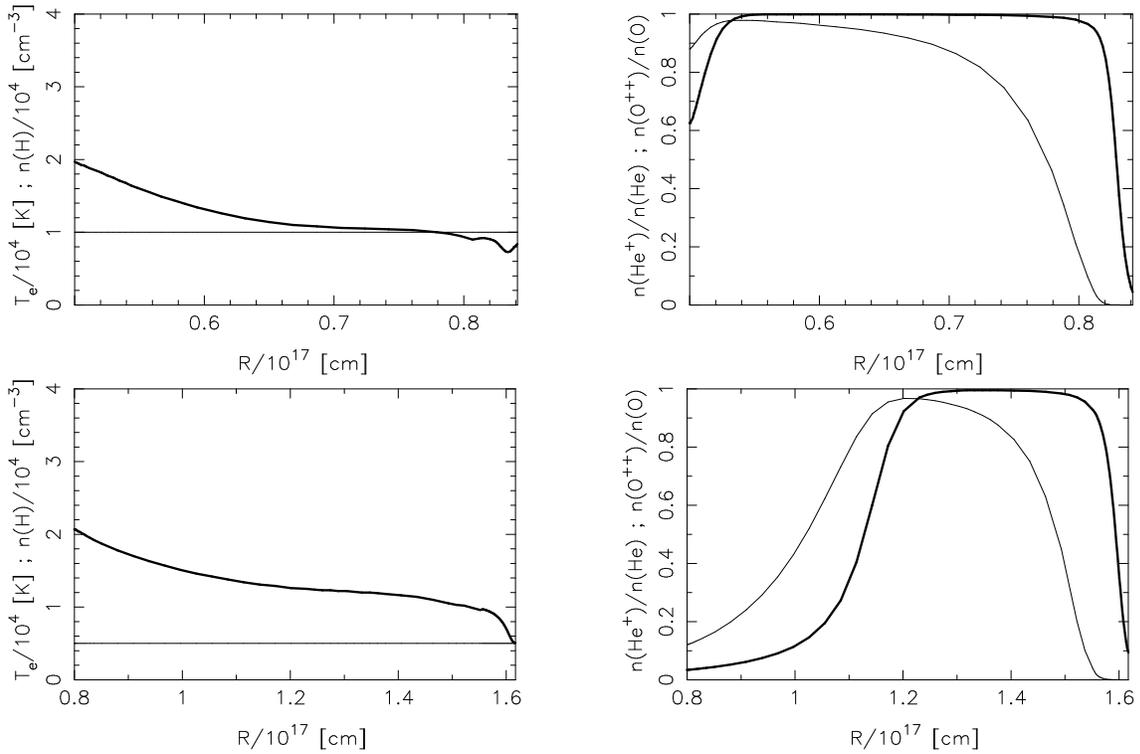

\centerline{\psfig{figure=MS1707f3a.ps,width=15.cm}}
\centerline{\psfig{figure=MS1707f3b.ps,width=15.cm}}
\caption{
The temperature and ionization structure of the same constant
density models as in Fig.\,\ref{Fig2} but in which there are
two populations of grains: small grains and large grains, each
population having a dust-to-hydrogen mass ratio of  5\,10$^{-3}$
(Models Ca and Cb). The presentation is the same as for the
models shown in Fig.\,\ref{Fig1}.
}
\label{Fig3}
\end{figure*}
Now, the \Hb\ fluxes are considerably reduced, by a factor 4 in the
$T_*$\,=\,10$^5$\,K model and by a factor 6 in the $T_*$\,=\,5\,10$^4$\,K
one, with respect to the dust-free models. Indeed, the mass absorption
efficiency of small graphite grains is much larger than that of large ones
(see Draine \& Laor\,\cite{draine93}). In addition, small grains are
heated to larger temperatures, because the heating of grains is proportional
to their surface while the cooling is proportional to their volume. This
translates into much larger fluxes in the 12\,$\mu$m band (by
a factor of about 4) and in the 25\,$\mu$m band (by a factor of about 2)
when comparing with models computed solely with the MRN dust, as seen
in Tables\,\ref{Table3} and \ref{Table4}.

What is striking from the comparison of Fig.\,\ref{Fig3} with
Figs.\,\ref{Fig1} and \ref{Fig2} is the very significant increase in the
gas temperature when a population of small grains is included.
Figure\,\ref{Fig4} shows the contribution of the gas heating due to the
photoelectric effect with respect to the total heating (thick line)
and the ratio between gas cooling due to grain-gas collisions and the total
cooling (thin line) as a function of the distance to the central star.
\begin{figure}
\centerline{\psfig{figure=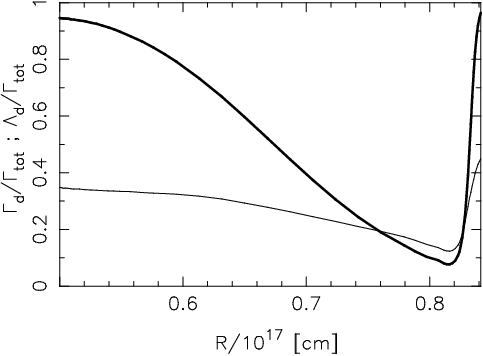,width=7.5cm}}
\centerline{\psfig{figure=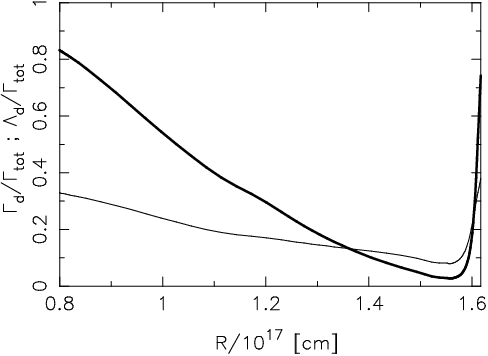,width=7.5cm}}
\caption{
The contribution of the heating due to the photoelectric effect from dust
grains with respect to the total heating (thick line) and the proportion of
cooling by grain-gas collisions with respect to the total cooling
(thin line) as a function of the distance to the
central star for the models shown in Fig. \,\ref{Fig3} (Models Ca - 
top and Cb - bottom).
}
\label{Fig4}
\end{figure}
As expected, the contribution of photoelectric heating is largest 
close to the star.
Indeed, the heating due to photoelectrons from dust grains per unit 
volume of gas,
$\Gamma_{\rm d}$, is proportional to the number density of dust 
grains and to the intensity
of the stellar radiation field. The heating due to photoionization of hydrogen,
$\Gamma_{\rm H}$, is proportional to the number density of neutral hydrogen
  and to the stellar radiation field. Using the ionization equilibrium equation
for hydrogen, one then finds that $\Gamma_{\rm H}$ is simply
proportional to the square of the gas density. It follows that the ratio
  $\Gamma_{\rm d}/\Gamma_{\rm H}$ is simply proportional to  $(m_{\rm 
d}/m_{\rm H})$ $U$,
where $U$ is the local ionization parameter,
defined as $Q({\rm H}^0)/(4\,\pi\,R^2\,n_{\rm e}\,{\rm c})$ (c is
the speed of light).
Therefore, the photoelectric effect becomes insignificant in fully 
ionized plasma
at large distances from the central stars. This results in an important
electron temperature gradient, which was also seen in the models of
Dopita \& Sutherland\,(\cite{dopita00}). Consequently,  \rOiii\ is 
increased, and the intensities of
collisionally excited lines such as \Oiii, \Oiiit\
or O\,{\sc{iii}}]\,$\lambda$1663 are increased with respect to \Hb, as seen in
Table\,\ref{Table3} and \ref{Table4}. As already noted by previous workers
(see Baldwin et al.\cite{baldwin91}) cooling due to gas-grain collisions can
also be important in the energy budget of the gas, and the net effect of the
grains on the gas temperature is not necessarily positive. In the models shown
in Fig.\,\ref{Fig4}, the net effect is slightly negative in the \Op\ zone.

Another noteworthy effect of the presence of small grains is the strong
heating of the gas due to the photoelectric effect on grains near the
ionization front. In this region, the Lyman continuum photons have
been exhausted, and the only photons that are present are
stellar and nebular photons below the Lyman limit (including the Ly
$\alpha$ radiation) which are not absorbed by hydrogen, but can heat
the gas via the photoelectric effect from dust grains. This results in
an enhancement of the [O\,{\sc{i}}]\,$\lambda$6300 line, as seen in 
Tables 3 and 4.

The heating efficiency of the small dust grains is such that even if
present in a very small amount, they can affect the gas
thermal balance significantly. We have run models Da and Db which are
identical to
models Ca and Cb except that the dust-to-hydrogen mass ratio of each grain
population is  5\,10$^{-4}$ instead of 5\,10$^{-3}$. Table\,\ref{Table3}
shows that, in the  $T_*$\,=\,5\,10$^4$\,K models, such a small amont of dust
increases the \rOiii\ ratio from 3.31\,10$^{-3}$ (dust free case) to
5.31\,10$^{-3}$ and the electron temperature $T_0$(\Opp)
from 8200 to 9200K.
Such a difference is quite significant if one is interested in an
accurate model fitting of a nebula. When comparing the models with
$T_*$\,=\,10$^5$\,K in Table 4 (models Ab and Db), the results are less
spectacular: the increase in $T_0$(\Opp) is only by 500\,K. This is because
the photoelectric heating contribution is important in the inner regions,
where oxygen is in the form of \Oppp\ and not of \Opp.
The effect of dust heating is better seen on lines of more charged ions,
  such as C\,{\sc{iv}}$\lambda$1550 whose intensity with respect to 
\Hb\ (not shown in
Table\,\ref{Table4}) is increased by a factor 2.

One way to measure a temperature gradient in a nebula is to compare
  $T_0$(\Opp) and $T_0$(\Op). We see from Table 3 and 4 that
models with small grains increase the $T_0$(\Opp)/$T_0$(\Op) ratio.
As a matter of fact, as already stated by Dopita \&
Sutherland\,(\cite{dopita00}),
dust could help explaining the observed temperature gradients in
planetary nebulae. Indeed, the ratio $T_{\rm r[O\,III]}$/$T_{\rm r[N\,II]}$
takes values between 1 and 1.5
in the planetary nebulae sample of Kingsburgh \&
Barlow\,(\cite{kingsburgh94}). Such gradients are very difficult to produce
with dust-free photoionization models. The most natural cause to
investigate is density gradients, since the temperature is
expected to be higher in regions of higher density, due to collisional
deexcitation of the cooling lines. Figure 5 displays such models
\begin{figure*}[!ht]
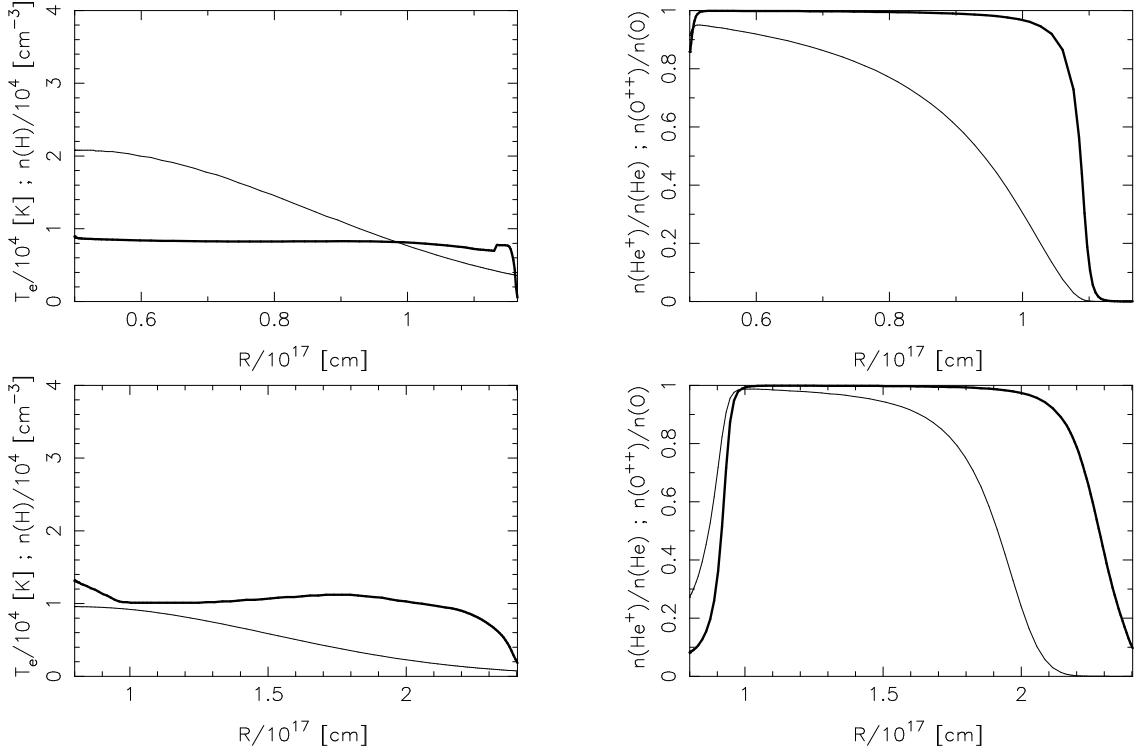

\centerline{\psfig{figure=MS1707f5a.ps,width=15.cm}}
\centerline{\psfig{figure=MS1707f5b.ps,width=15.cm}}
\caption{
The temperature and ionization structure of
dust-free models with a density gradient, models Ea and Eb (see text).
The presentation is the same as for the
models shown in Fig.\,\ref{Fig1}.
}
\label{Fig5}
\end{figure*}
(Ea and Eb), in which the density at given distance $R$ from the star
is given by:
\begin{equation}
n({\rm H})\,=\,n({\rm H})_{\rm in}\,{\rm 
exp}\left\{-\left(\frac{R\,-\,R_{\rm in}}{h}\right)^2\right\},
\end{equation}
where $R_{\rm in}$ is the inner radius of the nebula and $h$
the thickness scale of the envelope chosen to be 5\,10$^{16}$\,cm
for $T_*$\,=\,5\,10$^4$\,K and 10$^{17}$\,cm
for $T_*$\,=\,10$^5$\,K, respectively. These models have roughly the
same outer radius and \Hb\ luminosity as models Aa and Ab. Our 
computed models show that
the effect of such a density gradient is far from sufficient to
produce an appreciable temperature gradient. Of course, one can think
of much steeper density gradients, but in order to model a given
object, the gradient should be compatible with the observed surface  brightness
distribution and the density derived from \rSii\ and \rAriv. Our
experience is that it is generally very difficult to reproduce the
observed temperature gradient with a density gradient compatible
with the observations. Another possibility is to invoke depletion or
chemical abundance gradients. But then it is the outer parts of the
nebulae which should be richer in heavy elements, which makes such an
assumption rather unlikely. The hypothesis of photoelectric heating
by small dust grains provides a natural explanation to the outwards
decrease of electron temperatures that is inferred from the
observations. Of course, given all the unknowns as to the nature and
physical properties of such grains, a detailed model fitting
including small dust grains obviously would bear important
uncertainties.

\section{The effect of dust on the temperature profiles of planetary
nebulae with filamentary structures}

As mentioned in the introduction, density condensations have been
proposed to explain the temperature fluctuations inferred from the
differences between the temperatures derived from various indicators.
But in order to significantly enhance the temperature derived from
\rOiii, very high densities ($\sim 10^{6}$ cm$^{-3}$) are required.
Such densities are excluded
by observations (e.g. Liu et al. 2001a). Modest density fluctuations, though,
are obviously
present, as seen on numerous high resolution images of planetary nebulae
which reveal knots and filaments. In a dust-free nebula, the filaments
will have a lowered ionization compared to the ambient gas,
but the temperature will not be largely different from that of the 
ambient medium.
  We have modeled such a
situation by assuming that the filaments are concentric shells
and that the density distribution in the nebula is given by:

\begin{equation}
n({\rm H})\,=\,n({\rm H})_{{\rm min}} + \delta n({\rm H})\,\left({\rm 
cos}\frac{2 \pi R}{\delta r}\right)^{2q},
\end{equation}
where $q$ is taken equal to 15 in order to produce filaments that
occupy a small volume of the nebula.

\begin{figure*}[!ht]
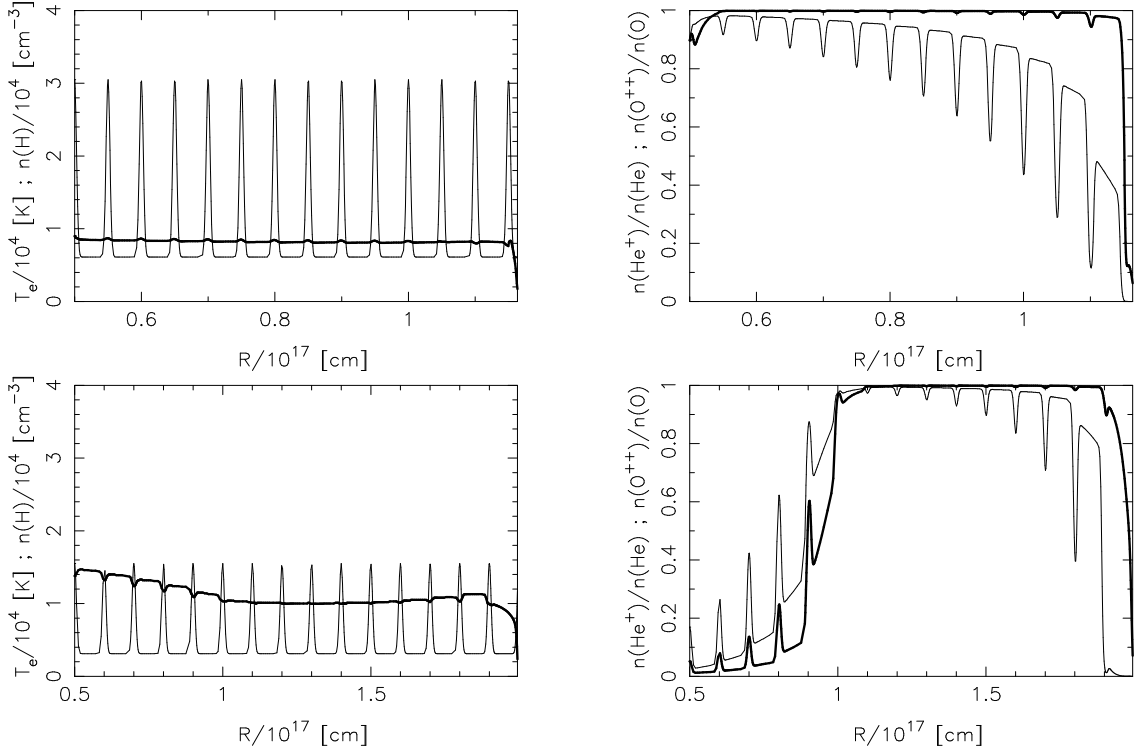

\centerline{\psfig{figure=MS1707f6a.ps,width=15.cm}}
\centerline{\psfig{figure=MS1707f6b.ps,width=15.cm}}
\caption{
The temperature and ionization structure of the dust-free filamentary models
Fa and Fb, which have
the same total \Hb\ flux as models Aa and Ab (see text).
Same presentation as in Fig.\,\ref{Fig1}.
}
\label{Fig6}
\end{figure*}
Figure\,\ref{Fig6} shows such
filamentary model nebulae (models Fa and Fb) whose density distribution
has been chosen to provide
the same \Hb\ luminosity and size as models Aa and Ab,
with a density contrast ${\delta}n({\rm H})$ of a factor 5.
Note that such completely ionized filaments are far from pressure
equilibrium with the ambient gas, so their lifetime is very
short (about 30 years for filaments of thickness 10$^{15}$cm) which
means that they continuously form and disappear during
the lifetime of the planetary nebula. If however the gas contains a
population of small dust grains, the temperature inside the filaments
is expected to be smaller than that of the ambient gas, since the
importance of gas heating due to the photoelectric effect is roughly
proportional to the ionization parameter, as stated above. Therefore, a
decrease of $U$ in the denser regions will result in smaller heating due to
the photoelectrons from dust grains. Figure\,\ref{Fig7} illustrates 
such a case,
\begin{figure*}[!ht]
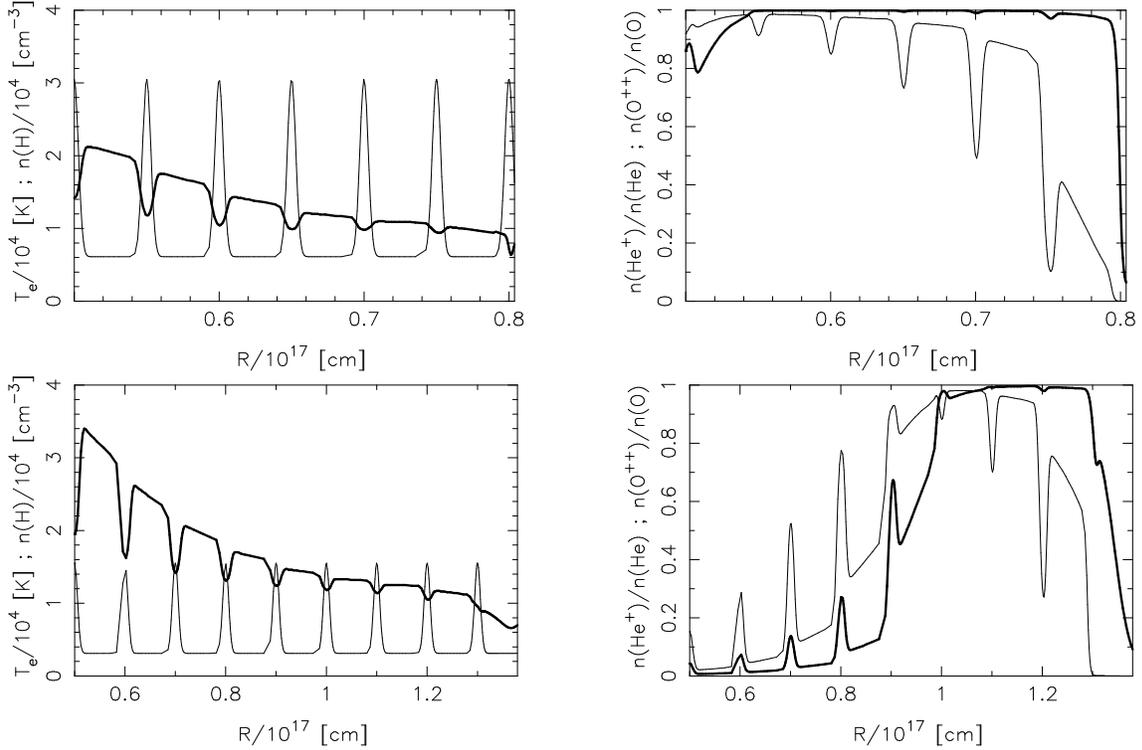

\centerline{\psfig{figure=MS1707f7a.ps,width=15.cm}}
\centerline{\psfig{figure=MS1707f7b.ps,width=15.cm}}
\caption{
The temperature and ionization structure of filamentary models (Ga and Gb)
with similar input parameters as models Fa and Fb except that they contain
dust with the same characteristics as models Ca and Cb.
Same presentation as in Fig.\,\ref{Fig1}.
}
\label{Fig7}
\end{figure*}
by showing models Ga and Gb, which have the same input values as models
Fa and Fb except that they contain dust with the same characteristics as models
Ca and Cb, respectively (i.e. small and large grains, each population with a
dust-to-hydrogen mass ratio of  5\,10$^{-3}$). Of course,
the \Hb\ luminosity and the size of the ionized region is smaller than in
models Fa and Fb, due to the fact that dust grains compete with hydrogen to
absorb the H Lyman continuum photons.  We clearly see the predicted
effect: the temperature inside the filaments is lower than in the
ambient medium. The magnitude of the difference decreases with the
distance to the ionizing star. Note  that in such a model, the
pressure discontinuities at the borders of the filaments
are much smaller than in a dust-free model. In other words, filaments in
a nebula containing small grains can survive longer than in a dust-free nebula.

\section{The effect of dust on the emission line analysis of planetary nebulae}

It is instructive to analyze our models with techniques used for the
interpretation of the spectra of real planetary nebulae. The virtue of such
"numerical observations" is that they eliminate the uncertainties in the
atomic data, since they can be done using exactly the same atomic data as
used in the computations of the models. Such an approach has been applied in
different contexts by Gruenwald \& Viegas\,(\cite{gruenwald98})
and Perinotto et al.\,(\cite{perinotto98}). Here we use our code
ABELION to analyze the models computed with PHOTO with the same atomic
data.

\subsection{Plasma diagnostics and abundance determinations from
integrated spectra}

We have used the usual procedure of deriving
$T_{\rm r[O\,III]}$ from \rOiii, $T_{\rm r[N\,II]}$
from \rNii,  $n_{\rm e[S\,II]}$ from \rSii\ and  $n_{\rm e[Ar\,IV]}$
from \rAriv\ ratio.
The emissivities of hydrogen and helium ions as well as those of
doubly ionized species of C, N, O and Ne were computed using $T_{\rm 
r[O\,III]}$
and $n_{\rm e[Ar\,IV]}$ while those of singly ionized species of C, N, O
and Ne
were computed using $T_{\rm r[N\,II]}$ and  $n_{\rm e[S\,II]}$.
The atomic abundance
ratios were then derived from the ionic abundance ratios using the ionization
correction factors of Kingsburgh \& Barlow\,(\cite{kingsburgh94}). The
abundance of O was obtained using the value of \Opp/\Hp\ as derived from
\Oiii/\Hb. The C/O, N/O and Ne/O ratios were derived from the \Ciii/\Oiii,
\Nii/\Oii\ and \Neiii/\Oiii, respectively.  The results
  are displayed in Tables\,\ref{Table5} and \ref{Table6},
respectively. For oxygen, we also show the determinations of
\begin{table*}
\caption{Abundances determination for models with $T_*$\,=\,5\,10$^4$\,K.}
\begin{flushleft}
\begin{tabular}{lccccccc}
\hline
    &
  model Aa &
  model Ba &
  model Ca &
  model Da &
  model Ea &
  model Fa &
  model Ga \\
\hline
$T_{\rm r[O\,III]}$              &  8242.    	& 8552. 
&13504.        & 9356.    &       8330.	   &      8422.	       & 
13145.   \\
$T_{\rm r[N\,II]}$                    &  8197.    	& 8434. 
&10480.	       & 9079.	  &       8776.	   &      9051.	       & 
11334.   \\
   &   &   &   &   &   &   \\
$n_{\rm e[S\,II]}$             &  9120.        & 8837.	        & 
6832.	       & 7444.    &       6548.	   &     14782.	       & 
10108.   \\
  $n_{\rm e[Ar\,IV]}$            & 11120.     	&11091. 
&10949.	       &11012.	  &      20401.	   &     13830.	       & 
12338.   \\
   &   &   &   &   &   &   \\
\Hep($\lambda$5876)/\Hp       & 1.11E-01   	&1.13E-01 
	&1.07E-01&	1.14E-01	&1.12E-01       &1.12E-01 
& 1.08E-01 \\
\Hepp($\lambda$4686)/\Hp      & 1.34E-04   	&1.88E-04 
	&7.36E-04&	2.03E-04	&1.37E-04	&1.35E-04 
	&6.18E-04 \\
He/H                          & 1.12E-01   	      & 1.14E-01 
	&1.08E-01&	1.15E-01	&1.12E-01	&1.12E-01 
	&1.08E-01 \\
   &   &   &   &   &   &   \\
\Op($\lambda$3727)/\Hp & 1.20E-04  	      & 1.12E-04 
	&7.04E-05&	8.23E-05	&6.34E-05	&8.83E-05 
	&5.55E-05 \\
\Opp($\lambda$5007)/\Hp& 3.26E-04   	      & 3.30E-04 
	&2.61E-04&	3.32E-04	&3.57E-04	&2.77E-04 
	&2.18E-04 \\
\Opp($\lambda$1663)/\Hp       & 3.37E-04	&3.42E-04 
	&3.08E-04&	3.56E-04	&3.71E-04	&2.73E-04 
	&2.63E-04 \\
\Opp($\lambda$52\,$\mu$m)/\Hp & 3.27E-04	&3.31E-04 
	&3.08E-04&	3.43E-04	&4.04E-04	&2.66E-04 
	&2.25E-04 \\
\Opp($\lambda$88\,$\mu$m)/\Hp & 3.26E-04	&3.30E-04 
	&3.05E-04&	3.42E-04	&4.17E-04	&2.83E-04 
	&2.35E-04 \\
\Opp($\lambda$4651)/\Hp       & 3.28E-04	&3.33E-04 
	&3.48E-04&	3.53E-04	&3.56E-04	&2.92E-04 
	&3.06E-04 \\
O/H                           & 4.47E-04	&4.43E-04 
	&3.32E-04&	4.15E-04	&4.21E-04	&3.66E-04 
	&2.75E-04 \\
   &   &   &   &   &   &   \\
C/O                           & 1.29E+00	&1.28E+00 
	&1.17E+00&	1.23E+00	&1.28E+00	&1.31E+00 
	&1.19E+00 \\
   &   &   &   &   &   &   \\
N/O                           & 4.14E-01	&4.20E-01 
	&5.21E-01&	4.70E-01	&5.23E-01	&5.67E-01 
	&6.86E-01 \\
   &   &   &   &   &   &   \\
Ne/O                          & 2.62E-01	&2.65E-01 
	&2.73E-01&	2.68E-01	&2.59E-01	&2.71E-01 
	&2.83E-01 \\
\hline
\end{tabular}
\end{flushleft}
\label{Table5}
\end{table*}
\begin{table*}
\caption{Abundances determination for models with $T_*$\,=\,10$^5$\,K.}
\begin{flushleft}
\begin{tabular}{lccccccc}
\hline
    &
  model Ab &
  model Bb &
  model Cb &
  model Db &
  model Eb &
  model Fb &
  model Gb \\
\hline
$T_{\rm r[O\,III]}$  & 10388.&  10588.&  12968.&   10974.& 10597.& 
10472.&   13067.\\
  $T_{\rm r[N\,II]}$ & 10303.&  10247.& 10770.&  10500.&   10988.& 
11159.&   11494.\\
   &   &   &   &   &   &   \\
$n_{\rm e[S\,II]}$             &  3947.	 &  3964.	  & 4105. 
	 &  3975.	&   2180.&	   5439.&	    6171.\\
  $n_{\rm e[Ar\,IV]}$            & 5649.	 &  5650.	  & 5706. 
	 &  5645.	&   8916.&	   8935.&	    9394.\\
\\
   &   &   &   &   &   &   \\
\Hep($\lambda$5876)/\Hp       & 1.07E-01&	 1.05E-01 
	&8.67E-02	&1.05E-01&	 1.05E-01&	1.08E-01& 
	 8.58E-02\\
\Hepp($\lambda$4686)/\Hp      & 8.73E-03	&1.10E-02 
	&2.69E-02	&1.13E-02&	 9.09E-03&	8.96E-03& 
	 2.82E-02\\
He/H                          & 1.16E-01	&1.16E-01 
	&1.14E-01	&1.16E-01&	 1.14E-01&	1.17E-01& 
	 1.14E-01\\
   &   &   &   &   &   &   \\
\Op($\lambda$3727)/\Hp     &     5.70E-05   &    6.13E-05   & 
7.34E-05   &   6.02E-05   &   3.09E-05  &  4.55E-05  &  6.20E-05 \\
\Opp($\lambda$5007)/\Hp    &  	3.79E-04	&3.64E-04 
	&2.86E-04	&3.65E-04	&4.08E-04&	3.51E-04& 
	 2.61E-04\\
\Opp($\lambda$1663)/\Hp       &      	3.93E-04&	 3.79E-04 
	&3.08E-04	&3.79E-04	&4.25E-04&	3.60E-04& 
	 2.83E-04\\
\Opp($\lambda$52\,$\mu$m)/\Hp &    	3.89E-04	&3.76E-04 
	&3.08E-04	&3.77E-04	&4.84E-04&	3.76E-04& 
	 3.01E-04\\
\Opp($\lambda$88\,$\mu$m)/\Hp &    	3.91E-04	&3.77E-04 
	&3.09E-04	&3.78E-04	&5.33E-04&	4.33E-04& 
	 3.52E-04\\
\Opp($\lambda$4651)/\Hp       &      	3.89E-04	&3.76E-04 
	&3.16E-04	&3.77E-04	&4.18E-04&	3.67E-04& 
	 2.94E-04\\
O/H                           &    	4.59E-04	&4.55E-04 
	&4.30E-04	&4.56E-04	&4.64E-04&	4.18E-04& 
	 3.91E-04\\
   &   &   &   &   &   &   \\
C/O                           &  	1.04E+00	&1.02E+00 
	&9.93E-01	&1.00E+00	&1.03E+00&	1.06E+00& 
	 9.97E-01\\
   &   &   &   &   &   &   \\
N/O                           &        	5.88E-01	&5.80E-01 
	&5.82E-01	&5.87E-01	&6.26E-01&	8.12E-01& 
	 7.55E-01\\
   &   &   &   &   &   &   \\
Ne/O                          &	3.01E-01&	 3.04E-01 
	&3.15E-01	&3.03E-01	&2.85E-01&	3.17E-01& 
	 3.29E-01 \\

\hline
\end{tabular}
\end{flushleft}
\label{Table6}
\end{table*}
\Opp/\Hp\ based on the other lines produced by \Opp\ (including the
recombination line  O\,{\sc{iii}}$\,\lambda$4651) but still using
$T_{\rm r[O\,III]}$). Since these lines have different dependences on the
electron temperature (and density), differences between
these various determinations of the \Opp\
abundance in the same model from integrated "numerical spectra"
are a good measure of the temperature variations in the models (a
somewhat similar approach was adopted by Mathis et al.\,\cite{mathis98})
in the discussion of observational data on planetary nebulae.

Inspection of Tables\,\ref{Table5} and \ref{Table6} reveals that the only
cases where the computed O/H is significantly different from the input value
are models Ca, Fa and Ga
\footnote{One does not expect the O/H determinations to be {\em exactly}
equal to the input values, since the temperature characteristic of the
emission of the \Hb\
line is not equal to  $T_{\rm r[O\,III]}$ and the correction for "unseen"
oxygen ions introduces a slight error.}.
In the constant density case with small grains (model Ca), O/H is
underestimated by about 40\%, because of the strong temperature gradient
due to photoelectric heating by dust. In the dust-free case with
filaments (case Fa), O/H is underestimated by 30\%,  mainly because
the determination of the \Op\ abundance is biased by the presence of clumps
of density higher than the critical density for the \Oii\ line. The case where
the difference between the derived O/H and the true one is the largest
(almost a factor 2) is in model Ga, where small grains are present in
a clumpy medium. Then, the temperature inhomogeneities are quite large
and produce a significant bias in the oxygen abundance determination.
That the effect is mainly due to temperature inhomogeneities can be
seen from the fact that \Opp($\lambda$1663)/\Opp($\lambda$5007) is 1.21 and
\Opp($\lambda$4651)/\Opp($\lambda$5007) is 1.40, whereas these ratios are
0.985 and 1.05, respectively, for the filamentary dust-free case.
Note that in this model, the Balmer jump temperature (obtained by calibrating
our theoretical Balmer jump index on isothermal photoionization models) is
10800\,K, significantly smaller than the temperature derived from \rOiii\ and
\rNii. As regards the models with $T_*$\,=\,10$^5$\,K, although
qualitatively the effect of dust heating is similar to the one
occuring in the models with  $T_*$\,=\,5\,10$^4$\,K (see Figs.\,\ref{Fig3} and
\ref{Fig7}), the error in the derived O/H is much smaller. The reason is that
photoelectric heating is strongest in the region populated by \Oppp\ ions,
while the classical oxygen abundance determination relies on lines emitted in
the \Opp\ and \Op\ zones.
For the model which is most affected by temperature inhomogeneities
(the filamentary model with dust, i.e. model Gb), the Balmer jump
temperature is 11500\,K. This is significantly lower than  $T_{\rm r[O\,III]}$
and than $T_0$(\Hp), as can be seen in Table\,\ref{Table6} and \ref{Table4}.
For such a nebula,
the effect of dust heating would have dramatic consequences on the
interpretation of ultraviolet lines from higher ionization species, such as
\Cppp, \Nppp\ and \Oppp.

\subsection{Interpretation of spatially resolved observations}

We have also computed the electron temperatures, densities and
abundances for simulated observations on lines of sight located at
different projected distances from the central star.
\begin{figure*}[!ht]
\centerline{\psfig{figure=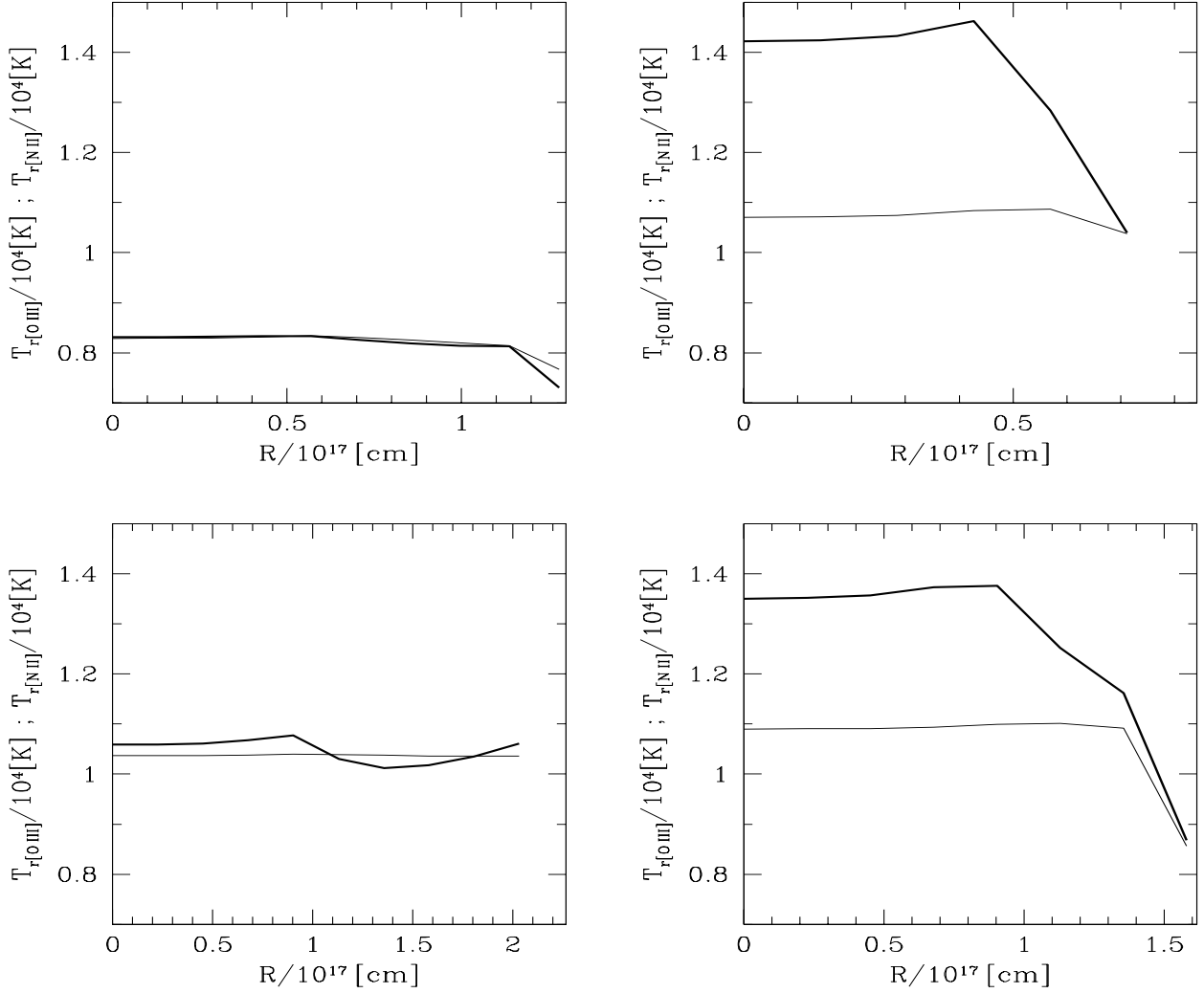,width=19.cm,height=16.cm}}
\caption{
$T_{\rm r[O\,III]}$ (thick lines) and $T_{\rm r[N\,II]}$ (thin lines)
  at different projected distances from the central star for constant density
dust-free models (Aa and Ab, left panels) and for constant density models with
two populations of grains: small grains and large grains, each
population having a dust-to-hydrogen mass ratio of  5\,10$^{-3}$
(Models Ca and Cb, right panels). As in the other figures, upper 
panels correspond to
    $T_*$\,=\,5\,10$^4$\,K and lower panels to $T_*$\,=\,10$^5$\,K.
}
\label{Fig8}
\end{figure*}
The temperature profiles $T_{\rm r[O\,III]}$ and $T_{\rm r[N\,II]}$ 
are shown in Fig.\,\ref{Fig8}
for our dust-free homogeneous models (Aa and Ab) and for the
homogeneous models with small grains (Ca and Cb).
We see that dusty models containing a population of small particles
(our models Ca and Cb) do show a significant gradient in $T_{\rm r[O\,III]}$
in contrast with dust-free models (Aa and Ab). For example $T_{\rm r[O\,III]}$
varies from 14200\,K (in the center) to 10500\,K (in the outskirts) in
model Ca, and from 13500\,K (in the center) to 8500\,K (in the outskirts) in
model Cb, while in dust-free models the  $T_{\rm r[O\,III]}$ variations are of
the order of 500\,K only. The variation of $T_{\rm r[N\,II]}$ across 
the face of
the nebulae is very mild (a few hundred K only), even in the case of
dusty models, which is not surprising since these lines arise in the
outer parts of the nebulae.

\section{The temperature fluctuation parameter}

Numerous studies have used the temperature
fluctuation concept introduced by Peimbert (1967) and derived $t^2$
from observational data (although as pointed out for example
by Kingdon \& Ferland 1995 the value derived from the observations
does not strictly correspond to the definition of  $t^2$).
Besides, as demonstrated by Stasi\'nska (2001) on two-zone toy 
models, different
temperature distributions with formally the same value of $t^2$
and same mean temperature can give very different ratios of
the various lines emitted by \Opp.

Because of this long tradition in the use of $t^2$, it is worth
commenting on the values of  $t^2$ found in our models
(they correspond to the  $t^{2}_{\rm str}$ values of Kingdon \& Ferland 1995).
As seen in Tables 3 and 4, and as noted by previous authors, the values of
$t^{2}$(\Hp), $t^{2}$(\Op) and $t^{2}$(\Opp) may actually be quite different.
They are smaller than 10$^{-2}$ for all our models
except those containing small dust grains with a dust-to-hydrogen
mass ratio  5\,10$^{-3}$. Models with only one population of grains
with the  MRN size distribution between 0.01 and 0.25 $\mu$m and a
dust-to-hydrogen mass ratio of 10$^{-2}$ (Models Ba and Bb) do give
higher values of $t^{2}$ than the analogous dust-free counterparts, but
they are still of the order of a few 10$^{-3}$ or less. In this case, actually,
higher values of $t^{2}$ result from a temperature gradient rather than
a small scale temperature variation. The dust-free models with 
inhomogeneous density
  distributions return values of $t^{2}$ that are similar to constant 
density models.

On the other hand, models containing small grains return values of 
$t^{2}$ largely above
10$^{-2}$. In the constant density cases, this is due to the strong
temperature gradients in the inner zone, which induce a  $t^{2}$(\Hp) of
  the order of 3\,-\,-4\,10$^{-2}$. For \Op, the value of  $t^{2}$ is smaller
  (but still around 10$^{-2}$)
because this zone is distant from the star and less affected by 
photoelectric heating.
A similar remark holds for $t^{2}$(\Opp) in the  $T_*$\,=\,10$^5$\,K model.

In the filamentary models with small grains (models Ga and Gb),
The values of $t^{2}$(\Hp) and $t^{2}$(\Opp) are further increased 
with respect to
the constant density models, because of the important small-scale temperature
variation seen in Fig. 7. This is the first time that photoionization models
give such large values of $t^{2}$ for nebulae with solar vicinity
chemical composition  and moderate effective
temperatures.

\section{Concluding remarks}

In continuation of previous studies on the effects of dust grains in
ionized plasmas (e.g. Spitzer\,\cite{spitzer48},
Baldwin et al.\,\cite{baldwin91},
Weingartner \& Draine\,\cite{weingartner01b}),
we have investigated the role of dust in the thermal
balance of planetary nebulae. This subject has been
introduced by the works of Borkowski \& Harrington\,(\cite{borkowski91}) and
Dopita \& Sutherland\,(\cite{dopita00}),
the first one studying the particular case of a
hydrogen-poor dust heated planetary nebula, the second one exploring
more generally the conditions under which dust heating plays a role 
in planetary
nebulae. These last authors found that dust heating is important
if a population of
small dust grains is assumed.

In the present study, we have analyzed in
more detail the consequences of the presence of such grains in
planetary nebulae, by considering several typical examples. In short, we
  have constructed photoionization models for two families of nebulae.
One concerns rather dense nebulae ($n$(H)\,=\,10$^4$\,cm$^{-3}$) surrounding
relatively cool central stars ($T_*$\,=\,5\,10$^4$\,K).
The other concerns more
diluted nebulae ($n$(H)\,=\,5\,10$^3$\,cm$^{-3}$) surrounding hot central
stars ($T_*$\,=\,10$^5$\,K).
Such cases provide good examples of different situations encountered in
planetary nebulae. We have studied the temperature profiles of such
models for different dust contents and different density
distributions. As expected, the effect of dust heating is
particularly important in the presence of a population of small
grains. The electron temperature increases by a factor of about 2 in the
central regions when adopting a population of small grains with a
dust-to-hydrogen mass ratio of 5\,10$^{-3}$.
Observational consequences of such a
temperature increase are seen in optical spectra only if the high
temperature zone contains a large proportion of \Opp\ ions which provide
good temperature diagnostics.

The presence of small grains in planetary nebulae can explain a
number of problems that have found no solution so far. For example,
tailored photoionization models of individual planetary nebulae
sometimes indicate an energy deficit in the thermal balance: the
electron temperature derived from the observation of  \rOiii\ is
higher than can be obtained by photoionization models. This is the
case in two of the five planetary nebulae with Wolf-Rayet central
stars studied by Pe\~na et al.\,(\cite{pena98}).

Also, high signal-to-noise, spatially
resolved observations of several
planetary nebulae show that the
temperatures derived from \rOiii\ smoothly decrease with projected
distance to the central star. Indirect indications of such gradients
are provided by ratios $T_{\rm r[O\,III]}$/$T_{\rm r[N\,II]}$ larger than one
in spatially resolved observations as well as in
integrated spectra of many planetary nebulae.
Such gradients are difficult to reproduce with
photoionization models, even considering density gradients.
They are naturally explained if one assumes that the nebulae contain
small grains.

Also, photoelectric heating by small dust grains naturally explains, at
least qualitatively, why in planetary nebulae Balmer jump temperatures
are smaller
than the temperatures derived from \rOiii.

Detailed photoionization modelling of planetary nebulae
often predicts too small a value of the
\Oi\ intensity as compared to observations
(Clegg et al. 1987, de Marco \& Crowther 1999, Dudziak et al. 2000).
One way to enhance this line in dust free plasmas is by assuming
high density optically thick condensations, which will strongly reduce
the local ionization parameter and  boost
the \Oi\ lines with respect to smooth density models.
However, \Oi\ can also be enhanced by additional heating in the transition zone
between ionized and neutral gas. This heating can be provided by
the photoelectric effect on small dust grains, whose importance
with respect to heating by photoionization increases as
the Lyman continuum photons
become exhausted.

Moderate density inhomogeneities, such as are inferred from the
distribution of the surface brightness of many planetary nebulae,
produce only minimal spatial temperature variations in dust-free nebulae.
The presence of small dust grains  completely modifies
the small scale temperature structure of filamentary planetary nebulae, by
spectacularly boosting the temperature
of the diluted component. This results in a situation closer to
pressure equilibrium between filament and ambient gas than in absence of
small grains.

It is perhaps by measuring directly  the temperature {\em in} and
{\em between} filaments that
one would have the best indirect indication of the presence of small
grains in planetary nebulae.

Such observations would be really worthwhile, since our models show 
that small dust grains
inside planetary nebulae are able to provide a natural
solution to the long standing problem of
temperature fluctuations first mentioned by Peimbert\,(\cite{peimbert67}).

There is still a huge amount of work to be done on dust properties
before photoelectric heating can be modelled to
the same degree of reliability as other heating processes in ionized
nebulae. Impressive progress in this sense has been made recently
(Weingartner \& Draine\,\cite{weingartner01a},
Li \& Draine\,\cite{li01}, Draine \& Li\,\cite{draine01}, Weingartner \&
Draine\,\cite{weingartner01b}) but many uncertainties remain.

We have shown that small dust grains are a promising explanation for
a number of problems encountered in planetary nebulae.
Still, several problems will not be solved. These are: i) the large
inverse temperature gradients materialized by the small
$T_{\rm r[O\,III]}$/$T_{\rm r[N\,II]}$
ratios observed in some planetary nebulae
(Pe\~na et al.\,\cite{pena01}); ii) the very low Balmer jump temperatures
(around 3.5\,10$^3$\,K) observed in a few planetary nebulae
(Liu et al.\,\cite{liu01b});
iii) the origin of the large discrepancies between
abundances derived from forbidden and collisionally excited lines
of the same ions.
Concerning the latter item however,
Liu et al. (2000) make the point that present computations of the
effective recombination coefficients for line emission
do not take into account dielectronic recombinations
for states with principal quantum number above 10. These processes 
are expected to
strongly enhance the emissivities of the recombination lines
at temperatures above \,2\,10$^4$\,K. Since small grains are able to boost
the electron temperature to such high values in the most tenuous parts
of filamentary or knotty planetary nebula,
they might well also solve the recombination line conundrum, if the speculation
about the enhancement of the recombination line emissivities at high 
temperatures
proves to be true.

\begin{acknowledgements}
This work was supported by the Jumelage France-Pologne, by the
Polonium program (contract N0 03242XJ) and the grant 2.P03D.020.17 from
the Polish Committee for Scientific Research. R.Sz. is grateful for the
hospitality of the DAEC in Paris-Meudon Observatory and G.S. for the
hospitality of NCAC in Toru\'n.
\end{acknowledgements}


\begin{thebibliography}{}

\bibitem[1979]{aitken79} Aitken, D.K., Roche, P.F., \& Spenser, P.M.
               1979, ApJ, 233, 925

\bibitem[1983]{aller83} Aller, L.H., \& Czyzak, S.J.
               1983, ApJS, 51, 211

\bibitem[1991]{baldwin91} Baldwin, J.A., Ferland, G.J., Martin, P.G., et al.
               1991, ApJ, 374, 580

\bibitem[1983]{barlow83} Barlow, M.J.
               1983, in Planetary Nebulae: IAU Symp. 103,
               ed. D.R. Flower (D. Reidel Publishing Company), 105

\bibitem[1993]{barlow93} Barlow, M.J.
               1993, in Planetary Nebulae: IAU Symp. 155,
               ed. R. Weinberger, A. Acker (Kluwer Academic Publishers), 163

\bibitem[1993]{binette93} Binette, L., Wang, J., Villar-Martin, M., et al.
               1993, ApJ, 414, 535

\bibitem[1991]{borkowski91} Borkowski, K.J. \& Harrington, J.P.
               1991, ApJ, 379, 168

\bibitem[1987]{Clegg87} Clegg, R.E.S., Harrington, J.P., Barlow, M.J., \&
                Walsh, J.R., 1987, MNRAS, 314, 551

\bibitem[1999]{cohen99} Cohen, M., Barlow, M.J., Sylvester, R.J., et al.
               1999, ApJ, 513, L135

\bibitem[1999]{deMarco99} de Marco, O.,  \& Crowther, P.A., 1999, 
MNRAS, 306, 931

\bibitem[1993]{draine93} Draine, B.T., \& Laor, A.
                1993, ApJ, 402, 441

\bibitem[2001]{draine01} Draine, B.T., \& Li, A.
                2001, ApJ, 551, 807

\bibitem[2000]{dopita00} Dopita, M.A., \& Sutherland, R.S.
               2000, ApJ, 539, 742

\bibitem[2000]{dudziak00} Dudziak, G., P\'equignot, D., Zijlstra, A.A.,  \&
Walsh, J.R., 2000, A\&A, 363, 717

\bibitem[2001]{esteban01} Esteban, C.
               2001, RevMexAA (Serie de Conferencias), in press

\bibitem[2001]{garnett01} Garnett, D.R.,  \& Dinerstein, H.L. 2001, 
ApJ, 558, 145

\bibitem[1996]{gesicki96} G{\c{e}}sicki, K., Acker, A.,  \& Szczerba, R.
               1996, A\&A, 309, 907

\bibitem[1973]{gillett73} Gillett, F.C., Forrest, W.J.,  \& Merrill, K.M.
               1973, ApJ, 183, 87

\bibitem[1998]{gruenwald98}  Gruenwald, R.,  \& Viegas, S.M.
                1998, ApJ, 501, 221

\bibitem[1988]{harrington88} Harrington, J.P., Monk, D.J,  \& Clegg, R.E.S.
               1988, MNRAS, 231, 577

\bibitem[1995]{kingdon95} Kingdon, J.B. \& Ferland, G.J.
               1995, ApJ, 450, 691

\bibitem[1994]{kingsburgh94} Kingsburgh, R.L.,  \& Barlow, M.J.
               1994, MNRAS, 271, 257

\bibitem[2001]{li01} Li, A.,  \& Draine, B.T.
               2001, ApJ, 550, 213

\bibitem[1993]{liu93} Liu, X.-W., \& Danziger, I.J.
               1993, MNRAS, 263, 256

\bibitem[2001b]{liu01a} Liu, X.-W., Barlow, M.J., Cohen, M., et al.
               2001a, MNRAS, 323, 343

\bibitem[2001a]{liu01b} Liu, X.-W., Luo, S.-G., Barlow, M.J., et al.
               2001b, MNRAS, in press

\bibitem[2000]{liu00} Liu, X.-W., Storey, P.J., Barlow, M.J., et al.
               2000, MNRAS, 312, 585


\bibitem[1989]{lenzuni89} Lenzuni, P., Natta, A.,  \& Panagia, N.
               1989, ApJ, 345, 306

\bibitem[1977]{mathis77} Mathis, J.S., Rumpl, W.,  \& Nordsieck, K.H.
               1977, ApJ, 217, 425

\bibitem[1998]{mathis98} Mathis, J.S., Torres-Peimbert, S.,  \& Peimbert, M.
                1998, ApJ, 495, 328

\bibitem[1981]{natta81} Natta, A., \& Panagia, N.
               1981, ApJ, 248, 189

\bibitem[1967]{peimbert67} Peimbert, M.
               1967, ApJ, 150, 825

\bibitem[1998]{pena98} Pe\~na, M., Stasi\'nska, G., Esteban, C. et al.
               1998, A\&A, 337, 866

\bibitem[2001]{pena01} Pe\~na, M., Stasi\'nska, G.,  \& Medina, S.,
               2001, A\&A, 367, 983

\bibitem[1980]{pequignot80} P\'equignot, D.,  \& Stasi\'nska, G.
               1980, A\&A, 81, 121

\bibitem[1998]{perinotto98} Perinotto, M., Kifonidis, K., 
Sch{\"o}nberner, R.,  \&
                Marten, H.
                1998, A\&A, 332, 1044

\bibitem[1984]{pottasch84} Pottasch S.R.
               1984, Planetary Nebulae (D. Reidel Publishing Company)

\bibitem[1989]{roche89} Roche, P.F.
               1989, in Planetary Nebulae: IAU Symp. 131,
               ed. S. Torres-Peimbert (Kluwer, Dordrecht), 117

\bibitem[1978]{shields78} Shields, G.A.
               1978, ApJ, 219, 559

\bibitem[1983]{shields83} Shields, G.A.
               1983, in Planetary Nebulae: IAU Symp. 103,
               ed. D.R. Flower (D. Reidel Publishing Company), 259

\bibitem[1948]{spitzer48} Spitzer, L.Jr.
                1948, ApJ, 107, 6

\bibitem[2001]{stasinska01} Stasi\'nska, G.
               2001, RevMexAA (Serie de Conferencias), in press

\bibitem[1996]{stasinska96} Stasi\'nska, G., \& Leitherer, C.
               1996, ApJS, 107, 661

\bibitem[1999]{stasinska99} Stasi\'nska, G., \& Szczerba, R.
               1999, A\&A, 352, 297

\bibitem[2001]{szczerba01} Szczerba, R., G\'orny, S.K., Stasi\'nska G.,
               et al. 2001, Ap\&SS, 275, 113

\bibitem[1994]{viegas94a} Viegas, S.M., \& Clegg, R.E.S.
               1994, MNRAS, 271, 993

\bibitem[1994]{viegas94} Viegas, S.M., \& Contini, M.
               1994, ApJ, 428, 113

\bibitem[1997]{volk97} Volk, K., Dinerstein, H.,  \& Sneden, C.
               1997 in Planetary Nebulae: IAU Symp. 180
               ed. H.J. Habing, J.G.L.M. Lamers (Kluwer Academic Publishers),
               284

\bibitem[1998]{waters98} Waters, L.B.F.M., Beintema, D.A., Zijlstra, A.A., et
                al. 1998, A\&A, 331, L61

\bibitem[2001a]{weingartner01a} Weingartner, J.C.,  \& Draine, B.T.,
               2001a, ApJ, 548, 296

\bibitem[2001b]{weingartner01b} Weingartner, J.C.,  \& Draine, B.T.,
               2001b, ApJS, 134, 263
\end{thebibliography}
\end{document}